\documentclass[fleqn,usenatbib]{mnras}

\usepackage{newtxtext,newtxmath}
\usepackage[T1]{fontenc}

\DeclareRobustCommand{\VAN}[3]{#2}
\let\VANthebibliography\thebibliography
\def\thebibliography{\DeclareRobustCommand{\VAN}[3]{##3}\VANthebibliography}

\usepackage{graphicx}	% Including figure files
\usepackage{amsmath}	% Advanced maths commands
\newcommand{\dtri}{$D_{\rm M33}=97\pm 6$~kpc}
\newcommand{\rharc}{$r_{\rm half}=0.67^{\prime~+0.2}_{-0.1}$}
\newcommand{\rhpc}{$r_{\rm half}=186^{+58}_{-32}$~pc}
\newcommand{\dist}{$D=916^{+65}_{-53}$~kpc}%{$D=962^{+32}_{-32}$~kpc}
\newcommand{\Mv}{$M_V=-6.0\pm0.3$}
\newcommand{\Mg}{$M_g=-5.7\pm0.3$}
\newcommand{\Mr}{$M_r=-6.2\pm0.3$}
\newcommand{\Lum}{$L=2.2^{+0.7}_{-0.5}\times10^4\,{\rm L}_\odot$}
\newcommand{\coords}{RA $=$ 01$:$21$:$40.6, Dec $=$ 26$:$23$:$28}
\newcommand{\pvii}{Pisc~VII}
\title[Pisces VII - A satellite of M33]{Pisces VII/Triangulum III - M33's second dwarf satellite galaxy}

\author[M. L. M. Collins et al.]{Michelle L. M. Collins,$^{1}$\thanks{E-mail: m.collins@surrey.ac.uk (MLMC)}
Noushin Karim,$^{1}$
David Martinez-Delgado,$^{2}$
 Matteo Monelli $^{3,4}$, \newauthor 
Erik J. Tollerud$^{5}$,Giuseppe Donatiello$^{6}$,  Mahdieh Navabi$^{1}$,
Emily Charles$^{1}$,
 Walter Boschin$^{3,4,7}$
\\
$^{1}$School of Maths and Physics, University of Surrey, Guildford, GU2 7XH, UK\\
$^{2}$Instituto de Astrof\'isica de Andaluc\'ia, CSIC, Glorieta de la Astronom\'\i a, E-18080, Granada, Spain \\
$^{3}$Instituto de Astrof\'isica de Canarias (IAC), Calle V\'ia L\'actea s/n, E-38205 La Laguna, Tenerife; Spain \\
$^{4}$Facultad de F\'isica, Universidad de La Laguna, Avda. Astrof\'isico Fco. S\'anchez s/n, 38200La Laguna, Tenerife, Spain. \\
$^{5}$ Space Telescope Science Institute,  3700 San Martin Drive, Baltimore, MD 21218, USA \\
$^{6}$UAI -- Unione Astrofili Italiani /P.I. Sezione Nazionale di Ricerca Profondo Cielo, 72024 Oria, Italy \\
$^{7}$Fundaci\'on G. Galilei - INAF (Telescopio Nazionale Galileo), Rambla J. A. Fern\'andez P\'erez 7, E-38712 Bre\~na Baja (La Palma), Spain
}

\date{Accepted XXX. Received YYY; in original form ZZZ}

\pubyear{2015}

\begin{document}
\label{firstpage}
\pagerange{\pageref{firstpage}--\pageref{lastpage}}
\maketitle

\begin{abstract}
Pisces VII/Triangulum III (\pvii) was discovered in the DESI Legacy Imaging Survey and was shown to be a Local Group dwarf galaxy with follow-up imaging from the 4-m Telescopio Nazionale Galileo. However, this imaging was unable to reach the horizontal branch of \pvii, preventing a precision distance measurement. The distance bound from the red giant branch population placed \pvii\ as either an isolated ultra-faint dwarf galaxy or the second known satellite galaxy of Triangulum (M33). Using deep imaging from Gemini GMOS-N, we have resolved the horizontal branch of \pvii, and measure a distance of \dist, making \pvii\ a likely satellite of M33. We also remeasure its size and luminosity from this deeper data, finding \rhpc, \Mv\ and \Lum. Given its position in the M33 halo, we argue that \pvii\ could support the theory that M33 is on its first infall to the Andromeda system. We also discuss the presence of blue plume and helium burning stars in the colour-magnitude diagram of \pvii\ that are consistent with ages of $\sim1.5$~Gyr. If these are truly members of the galaxy, it would transform our understanding of how reionisation affects the faintest galaxies. Future deep imaging and dynamics could allow significant insight into both the stellar populations of \pvii\ and the evolution of M33. %We discuss the significance of this finding and the presence of potentially young blue stars in the colour magnitude diagram.
\end{abstract}


%%%%%%%%%%%%%%%%%%%%%%%%%%%%%%%%%%%%%%%%%%%%%%%%%%

%%%%%%%%%%%%%%%%% BODY OF PAPER %%%%%%%%%%%%%%%%%%

\section{Introduction}

With a stellar mass of $3\times10^9\,{\rm M_\odot}$ \citep{mcconnachie12} and a halo mass of $\sim10^{11}\,{\rm M_\odot}$ \citep{corbelli14}, the Triangulum galaxy (M33) is the third most massive system in the Local Group. Its recent orbital history is something that has been debated in the literature over the past decade. M33 has long been known to harbour a warped HI gas disk \cite[e.g.][]{rogstad76}, and deep imaging from the Pan-Andromeda Archaeological Survey (PAndAS) showed significant warping in its stellar disk and a potential stream between M31 and M33 \citep{mcconnachie09,mcconnachie10}. Both features have been attributed to a previous close passage (50-100~kpc) with M31 within the past $\sim3$~Gyr \cite[e.g][]{mcconnachie09,putman09,teppergarcia20}. However, other studies using a combination of M33's proper motion and inference from cosmological simulations suggest it is far more likely that M33 is on its first infall to the Andromeda system \cite[e.g.][]{patel17,vandermarel19}. 

These vastly different hypotheses carry significant implications for understanding our nearest neighbour galaxies, for example, their expected satellite systems. Any close passage between M33 and the vastly more massive M31 would have stripped much of M33's dark and stellar halo, as well as a number of its satellites. This naturally reduces the number of dwarf galaxies we expect to find around M33, which is currently satellite-poor compared to other Local Group spirals. To date, it has only one (mostly) secure satellite candidate -- Andromeda (And) XXII \citep{martin09,chapman13} -- which has a luminosity of $L=3.1\times10^4\,{\rm L_\odot}$ \citep{martin16,savino22}. Yet predictions from cosmological simulations suggest it should harbour $\sim8-10$ dwarf galaxies with $L>10^4\,{\rm L_\odot}$ \citep{dooley17,patel18}. If M33 did experience a close flyby with M31, this could explain its ``missing'' satellite issue. 

In the first infall scenario, we must instead assume that these satellites exist, but have not been found yet as they lie beneath the detection thresholds of current surveys. This is certainly plausible as the only deep, wide-field survey of M33 is PAndAS \citep{mcconnachie09,mcconnachie18}. This survey only probed M33 out to a radius of $\sim50$~kpc, which is about a third of its predicted virial radius \cite[e.g.][]{patel18}, and is only sensitive to satellites with $L\gtrsim2\times10^4\,{\rm L_\odot}$ \citep{dolivadolinsky22,dolivadolinsky23}. Thus, there may be a number of dwarf galaxies waiting to be uncovered both outside and within the PAndAS boundary \citep{patel18}. We cannot know for sure until we have a full, wide and deep survey of the M33 halo.

%The impact of a prior close fly-by with M31 goes beyond the properties of M33's satellite system. If it truly has punched through the M31 system, it will also impact our ability to measure meaningful dynamical properties for Andromeda, including its halo mass and radial profile. Numerous works studying the similarly massive Large Magellanic Cloud (which is currently on its first infall to the Milky Way, e.g. \citealt{besla07}) have shown that its presence is substantially affecting the inferred properties of the Galactic potential, and the orbits of its stellar streams and satellites \cite[e.g][]{veraciro13,gomez15,garavitocamargo19,garavitocamargo21,peterson20,erkal19,erkal20a,erkal21a,deason21,vasiliev21,lilleengen23,makarov23}. A close fly-by between M31 and M33 would then have a significant impact on studies aiming to constrain the mass of M31 using the dynamics of its satellites or stellar halo.

%Given this uncertainty in evolution, it is imperative that we constrain the orbital history of the Andromeda-Triangulum system. One way to do this is simply by searching for its dwarf satellite galaxies and stellar halo at large radii. If a number of bound satellites exist around M33 out to the full extent of its (undisturbed) stellar halo, the probability of a close passage is low. However if we find no trace of satellites or stellar halo beyond the confines of the PAndAS survey, this would support a close passage scenario.

With this in mind, we have been performing a systematic visual survey of shallow DESI Legacy Imaging Survey (DESI-LIS) data to hunt for faint dwarf galaxies that are too low surface brightness to be detected with automated routines like match-filter \cite[e.g.][]{drlica15,koposov15,cerny21a,cerny21b,dolivadolinsky22}. Our visual inspection
survey is complementary to these efforts, as our approach is more sensitive to partially resolved
systems which could be missed by the automatic detection algorithms. To date, we have found one M31 dwarf galaxy candidate \cite[Pegasus~V,][]{collins22a} and one system that is either an isolated Local Group ultra-faint dwarf galaxy (UFD) or the second known satellite of M33 -- Pisces VII/Triangulum III  \cite[\pvii,][]{martinezdelgado22}. \pvii\ was first identified as an over-density of partially resolved sources in DESI-LIS and followed up with the 4-m Telescopio Nazionale Galileo (TNG). Unfortunately, \citet{martinezdelgado22} could not well-constrain its distance. 

Our initial imaging also showed that \pvii\ may host stars as young as $\sim1$~Gyr based on an over-density of blue stars in its color magnitude diagram. This would be surprising for such a faint ($M_V=-6.1\pm0.2$, \citealt{martinezdelgado22}) dwarf galaxy, as such systems with stellar masses of order $\sim10^5\,{\rm M}_\odot$ are typically presumed to cease star formation shortly after the epoch of reionisation \cite[e.g.][]{bovill11a,bovill11b,brown14,wheeler15,rodriguezwimberly19}. However, recent observational studies have identified a number of ultra-faint dwarf galaxies with extended star formation. These include the isolated dwarf Pegasus~W \citep{mcquinn23} and the Andromeda satellite And~XIII \citep{savino23}. This may imply our understanding of the impact of reionisation on star formation in low mass galaxies is incomplete, or that these systems have reignited their star formation after reionisation by either retaining or accreting cold gas \cite[e.g.][]{rey20,applebaum21,gutcke22}.

To determine the distance to \pvii\ and to further shed light on the potential young stellar population in this system, we have acquired deeper Gemini/GMOS-N imaging of \pvii\ that allows us to identify the horizontal branch (HB) of this galaxy. In \S~\ref{sec:data}, we present the details of our imaging before deriving \pvii's structural parameters and distance in \S~\ref{sec:results}. We discuss the implications of our findings in \S~\ref{sec:discussion} before concluding in \S~\ref{sec:conclusions}. 

\section{Gemini GMOS-N imaging}
\label{sec:data}

To determine the distance to \pvii, we applied for follow-up imaging with the Gemini North GMOS-N instrument (Program ID GN-2022B-Q-132, PI Collins). \pvii\ was observed on 2022-08-08 in the $g-$ and $r-$band, split into $4\times450$~s and $4\times240$~s exposures respectively. The images were reduced using the Gemini {\sc Dragons} package to perform dark, bias and flat fielding, before stacking the resulting images into our final science frames. We next performed photometry using DAOPHOT/ALLFRAME \citep{stetson87,stetson94} in largely the same manner as \citet{monelli10b}, \citet{martinezdelgado22} and \citet{collins22a}. We identified stellar sources on each stacked image and performed aperture photometry, PSF derivation and PSF photometry with ALLSTAR.
We passed the final list of stars to ALLFRAME to construct catalogues in both $g-$ and $r-$band before combining the two. Finally, we calibrated our instrumental magnitudes by matching sources to photometry from the Sloan Digital Sky Survey \citep{abazajian09} using  linear relation in each band. We then use the data from \citet{schlafly11} to extinction correct our data in each band, where $A_g=0.357$ and $A_r=0.247$.

Star-galaxy separation was performed using the sharpness parameter in each band (requiring $-0.5<{\rm sharp}<+0.5$). In Fig.~\ref{fig:cmd} we present the colour-magnitude diagram (CMD) in the left-hand panel for the central $1.4^\prime$, which corresponds to two times the half-light radius (see \S~\ref{sec:structural}). The right-hand panel shows an equal area annulus located beyond three half-light radii for the candidate. A sparse red giant branch (RGB) is seen in CMD for the source, as well as a horizontal branch (HB) feature at $g\sim25.5$. Corresponding features are not seen in the field CMD.

%\subsection{Gemini GMOS-N photometry}

\section{Results }
\label{sec:results} % used for referring to this section from elsewhere
In this section, we present the structural parameters, distance and luminosity for \pvii. We summarise these findings in Table~\ref{tab:properties}. Our data analysis was conducted using Python, and relies heavily on packages including {\sc astropy} \citep{astropy:2013,astropy:2018,astropy:2022}; {\sc emcee} \citep{emcee}, {\sc corner} \citep{corner},{\sc NumPy} \citep{numpy}, {\sc SciPy} \citep{scipy} and {\sc matplotlib} \citep{hunter07}.

\subsection{Re-measuring the structural properties of \pvii}
\label{sec:structural}

We measured structural properties and luminosities for \pvii\ from our shallower TNG data. However, our deeper Gemini data allow us to update and refine these values. We follow the same method as presented in \citet{martinezdelgado22} and \citet{collins22a}. This method is based on the approach of \citet{martin16}, where we assume that our Gemini observations contain a dwarf galaxy candidate with $N_*$ observed members whose surface brightness profile can be modelled as an exponentially declining function with half-light radius $r_{\rm half}$ and an ellipticity $\epsilon$:

\begin{equation}
   \rho_{\rm dwarf}(r)=\frac{1.68^2}{2\pi r_{\rm half}^2(1-\epsilon)}N^*\exp{\left(\frac{-1.68r}{r_{\rm half}}\right)}
    \label{eq:density profile}
\end{equation}

\noindent where $r$ is the elliptical radius such that: 

\begin{equation}
    \begin{split}
        r= \Bigg( \Big(\frac{1}{1-\epsilon}((x-x_{0})\cos{\theta}-(y-y_{0})\sin{\theta})\Big)^2 \\
            +\Big((x-x_{0})\sin{\theta}-(y-y_{0})\cos{\theta}\Big)^2 \Bigg) ^{\frac{1}{2}}.
    \label{eq:radial profile}
    \end{split}
\end{equation}

\noindent Here, $\theta$ is the position angle of the major axis, $x_{0}$ and $y_{0}$ are the central co-ordinates of the dwarf galaxy and $x$ and $y$ are the coordinates on the tangent plane to the sky at the center of the Gemini GMOS-N field. We then assume that it is located in a field with a contaminant population that is uniform in density, described by $\Sigma_b$. We can determine $\Sigma_b$ by subtracting the total number of  dwarf galaxy stars (determined from the integral of the surface density profile in Eq.~\ref{eq:density profile}) from the total number of potential member stars within the field. We combine these assumptions into a likelihood function:

\begin{equation}
    \rho_{\rm model}(r) = \rho_{\rm dwarf}(r) + \Sigma_{b} .
    \label{eq:total profile}
\end{equation}

\noindent which we can sample using a Markov Chain Montel Carlo (MCMC) approach. As in prior work, we use the {\sc emcee} sampler \citep{emcee} to sample the posterior. We use limited uniform priors, requiring only that the number of stars within the dwarf galaxy $N_*\geq0$, the position angle falls in the range $0<\theta<\pi$ and the ellipticity lies between $0<\epsilon<1$.

\begin{figure*}
	\includegraphics[width=6in]{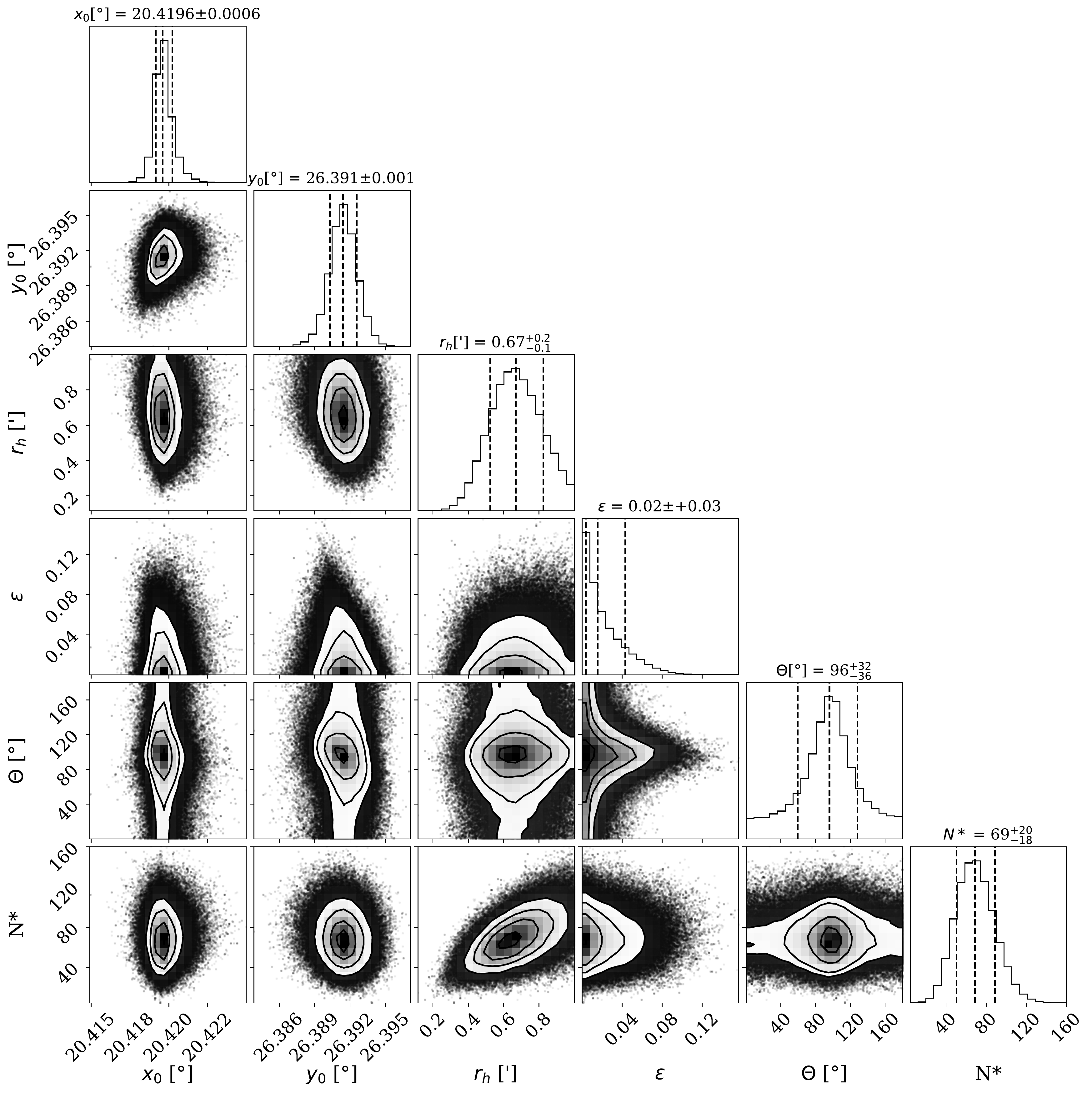}
	\caption{Here we show the posteriors for the structural properties of \pvii, derived using {\sc emcee} \citep{emcee} All properties other than the ellipticity ($\epsilon$) are well-constrained. These values are all consistent with our previous results presented in \citet{martinezdelgado22} within $1\sigma$. The precision has improved.  }
    \label{fig:corner}
\end{figure*}

Before running {\sc emcee}, we apply cuts to our data to ensure we only include sources down to our approximate completeness limit and to well-sample the contaminating population. We set a wide colour cut of $-0.2<g-r<2$, and only include sources with $r<26$-mag. We have modified both these cuts and find that our solutions are stable to shallower and narrower cuts. We include all sources within 2.5$^\prime$ of the centre.  We initialise {\sc emcee} with 100 walkers, and run for 10,000 iterations with a burn in of 3,250. These numbers were determined based on a visual inspection of the traces. We show the results in Fig.~\ref{fig:corner} and summarise them in Table \ref{tab:properties}. The solution is well-constrained, localising an over-density at \coords, with \rharc. There are not enough sources to reliably constrain the ellipticity, so we can only place a limit based on the 84\% confidence interval, giving $\epsilon<0.1$. Our results are entirely consistent with -- but more precise than -- the results of \citet{martinezdelgado22}.

\begin{figure}
	\includegraphics[width=\columnwidth]{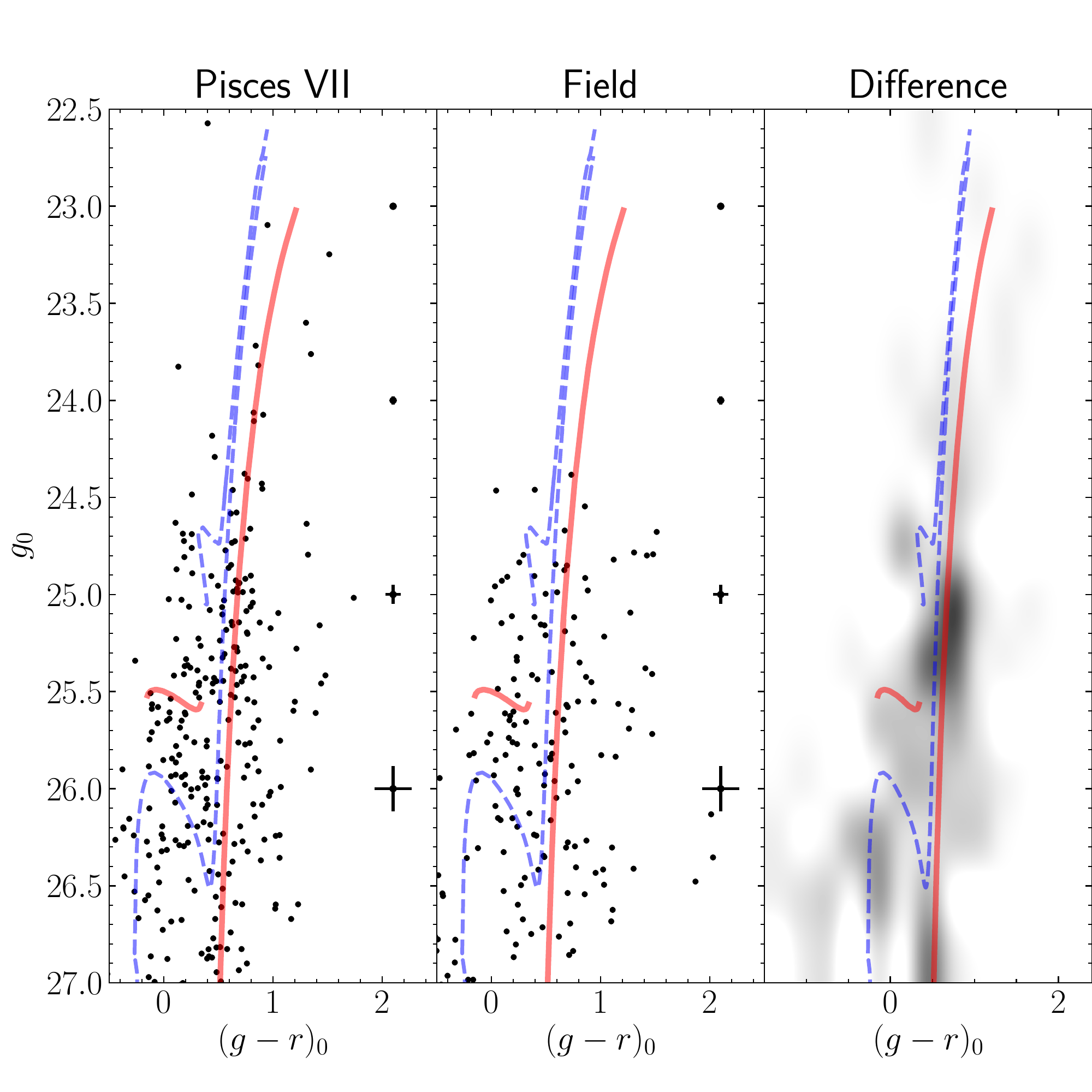}
	\caption{Here we present the deep Gemini-GMOS-N color magnitude diagram for \pvii. On the left, we show all stellar sources within 2$\times r_{\rm half}$, and in the centre we show an equal area annulus beyond 3$\times r_{\rm half}$. Finally, thr right panel shows a Hess diagram where we have subtracted the field CMD from the source CMD. The RGB of \pvii\ can be identified, as well as a horizontal branch feature at $g\sim25.5$~mag. We overlay two isochrones from the BaSTI library \citep{hidalgo18} with a metallicity of [Fe/H]$=-1.9$~dex, an alpha abundance of $[\alpha/{\rm Fe}]=+0.4$ and ages of 12.5 Gyr (red solid line) and 1.5 Gyr (blue dashed line). Both isochrones have been shifted to match the CMD by assuming a distance of 952~kpc. Intriguingly, the CMD shows an abundance of blue stars that are well traced by the younger isochrone in \pvii\ which are not seen in the field comparison, as well as potential helium-burning stars above the HB of \pvii. {\it Such young stars in an ultra-faint dwarf are completely unexpected, and if validated would have implications for our understanding of reionisation.}}
    \label{fig:cmd}
\end{figure}

\begin{table}
\centering
	\caption{The final structural and photometric properties for \pvii.}
	\label{tab:properties}
\begin{tabular}{cc}
\hline\\
Property & Value \\
\hline\\
RA ($^\circ$) & 20.419$\pm0.000$ \\
dec ($^\circ$) & 26.391$\pm0.001$ \\
$r_{\rm half} (^\prime$) & $0.67^{+0.2}_{-0.1}$ \\
$r_{\rm half}$ (pc) & $186^{+58}_{-32}$ \\
$\epsilon$ & $<0.1$ \\
PA ($^\circ$) & 96$^{+32}_{-36}$ \\
$N_*$ & 69$^{+20}_{-18}$ \\
(m-M)$_0$ & 24.8$\pm0.2$ \\
$D$ (kpc) & 916$^{+65}_{-53}$ \\
$D_{\rm M33 (kpc)}$ & 97$\pm6$ \\
M$_g$ & -5.7$\pm0.3$ \\
M$r$ & -6.2$\pm0.3$ \\
M$_V$ & -6.0$\pm0.3$ \\
$L ({\rm L_\odot})$ & 2.2$^{+0.7}_{-0.5}\times10^4$ \\
\hline
\end{tabular}
\end{table}

\subsection{The distance to \pvii}
\label{sec:dist}

In our prior work, we tried to constrain the distance to \pvii\ using its RGB stars. Given the paucity of bright stars in this galaxy, we were not able to make a precise measurement, finding that \pvii\ was located at a distance of $\sim0.8-1$~Mpc. However, our deeper data resolves a HB feature for \pvii\ at $g\sim25.5$-mag (see Fig.~\ref{fig:cmd}) which we can use as a distance indicator. First, we convert our SDSS photometry into the Johnson $BVRI$ magnitude system using the colour transformations of \citet{jordi06}, as the HB in the $V-$band has a near constant luminosity. Thus, by measuring the HB magnitude in the $V-$band, we can constrain the distance to \pvii.

We employ an MCMC approach to isolate the apparent magnitude of the HB in the $V-$band by fitting its luminosity function. Similar to the procedure of \citet{weisz19,collins22a} and \citet{mcquinn23}, we construct a likelihood function to capture the shape of the luminosity function such that:

\begin{equation}
    \mathcal{L}=e^{A(V_i-V_{\rm HB})}+e^{-0.5\left[\left(V_i-V_{\rm HB}\right)/C\right]^2}
\end{equation}

\noindent where $V_i$ is the $V-$band magnitude of the $i$-th star and $V_{\rm HB}$ is the best fitting magnitude for the HB.  $A, B$ and $C$ are free parameters. We use flat priors for all three, requiring $0 < A, B, C <2$. We initialise {\sc emcee} with 100 walkers, and run for 6,000 iterations with a burn in of 3,000. When marginalising over our free parameters, we measure $V_{\rm HB}=25.3\pm0.1$. 

To convert this into a distance measure, we calculate the expected absolute magnitude of the HB using the metallicity dependent calibration of \citet{carretta00}, where

\begin{equation}
    M_{V, {\rm HB}}= (0.13 \pm 0.09) * ([{\rm Fe/H}] +1.5)+(0.54\pm0.07).
\end{equation}

We use a value of [Fe/H]$=-1.9\pm0.2$~dex based on our best fit isochrone to the \pvii\ CMD, where the uncertainty is based on inspecting isochrones of higher and lower metallicity against the data. These values give $M_{V, {\rm HB}}=0.49^{+0.08}_{-0.04}$. Combining this with our measured apparent magnitude of $V_{\rm HB}=25.3\pm0.1$ we calculate a heliocentric distance for \pvii\ of \dist.

Now that we have 3D coordinates for \pvii, we can also place it in the wider context of the Andromeda-Triangulum group. Using the coordinates and distance for M31 and M33 from \citet{savino22} ($D=859^{+24}_{-23}$~kpc and $859^{+24}_{-23}$~kpc respectively) we measure these as $D_{\rm M31}=285^{+21}_{-18}$~kpc and \dtri\ for M31 and M33. If we assume the virial radius of M31 is $\sim250-300$~kpc, \cite[e.g.][]{patel18a,kafle18,blana18} and  M33 is $r_{200}\sim160$~kpc \citep{patel18}, we see that \pvii\ lies comfortably within the halo of M33, and in the outskirts of Andromeda's. It is thus a likely satellite of M33. We discuss this further in \S~\ref{sec:discussion}.

\begin{figure}
	\includegraphics[width=\columnwidth]{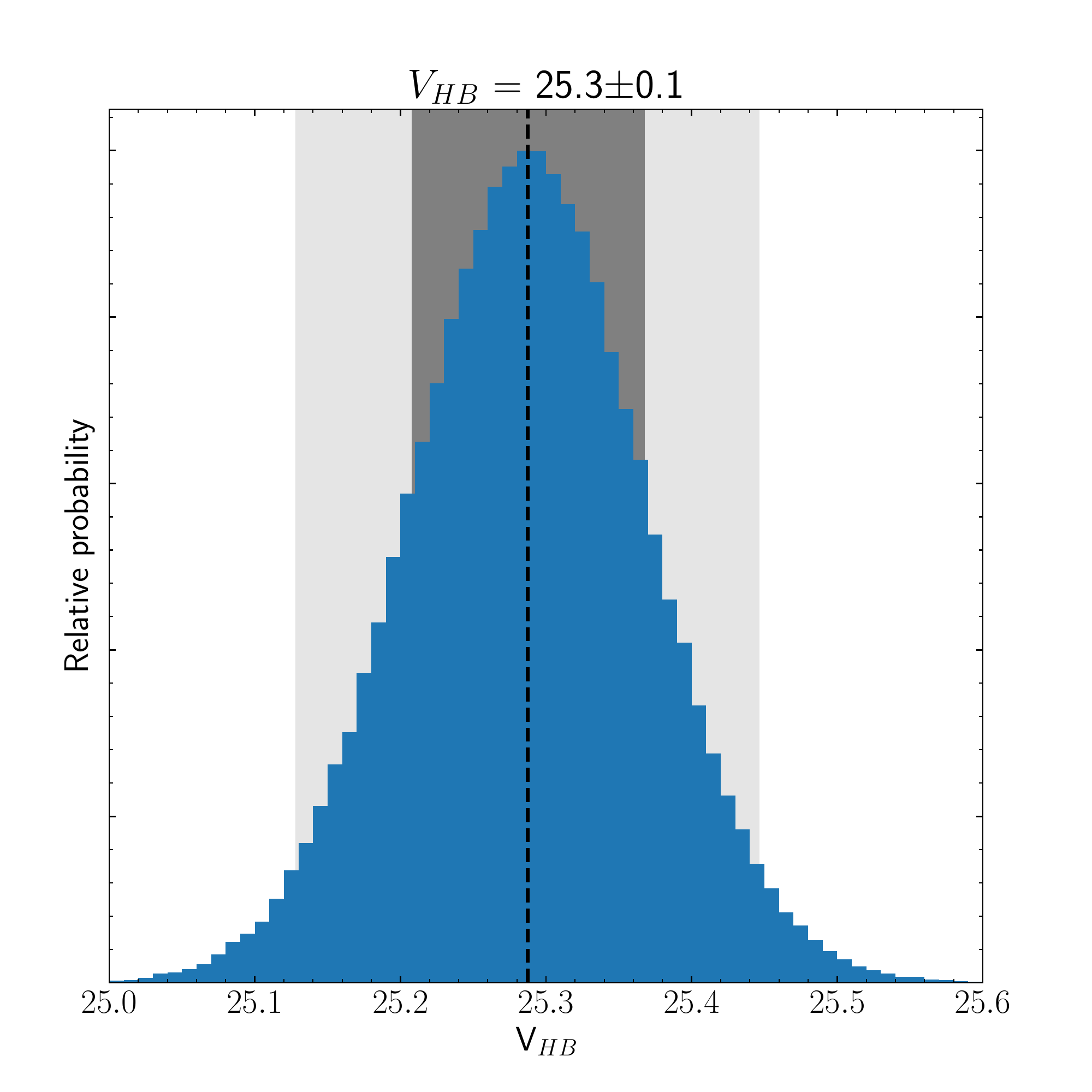}
	\caption{ This figure shows the marginalised probability distribution for the apparent magnitude of the horizontal branch of \pvii, $g_{\rm HB}$. The dashed line shows the median value, while the shaded gray regions show the $1$ and $2\sigma$ uncertainties. }
    \label{fig:hb}
\end{figure}

\subsection{The luminosity of \pvii}
\label{sec:luminosity}

Armed with a distance, we can now place a tighter constraint on the total luminosity of \pvii. In \citet{martinezdelgado22} we measured this to range from $-6.1<{\rm M}_V<-7.0$, depending on its distance. With the improved precision on our distance, we can reevaluate this with our deeper data. We follow the same method as \citet{martinezdelgado22} and \citet{collins22a}, which is again based on the procedure outlined in \citet{martin16}. We begin by downloading the luminosity function and corresponding isochrone for an old (age of 12.5 Gyr), metal-poor ([Fe/H]$=-1.9$) stellar population from the PARSEC database. We use the luminosity function as a probability distribution function (PDF) for the stellar population of \pvii. We shift the isochrone to the presumed distance of \pvii\ and use the PDF to perform a probability-weighted random selection of a magnitude in the $g-$band, and its corresponding $r-$band magnitude. We continue to select magnitudes until we have collected $N_*$ values that fall within the colour cuts we used to measure the structural properties of \pvii\ (see \S~\ref{sec:structural} for details). Once we have recorded $N_*$ values in this region, we then sum the magnitudes for all stars selected both within and outside of this cut to obtain final magnitudes in the $g$ and $r$-band. To account for shot noise and uncertainties in $N_*$ and distance, we repeat this procedure 1000 times, selecting a different value for distance and $N_*$ each time by randomly drawing from a Gaussian centered on our best fit values, with a width equal to their uncertainties.

\begin{figure}
	\includegraphics[width=\columnwidth]{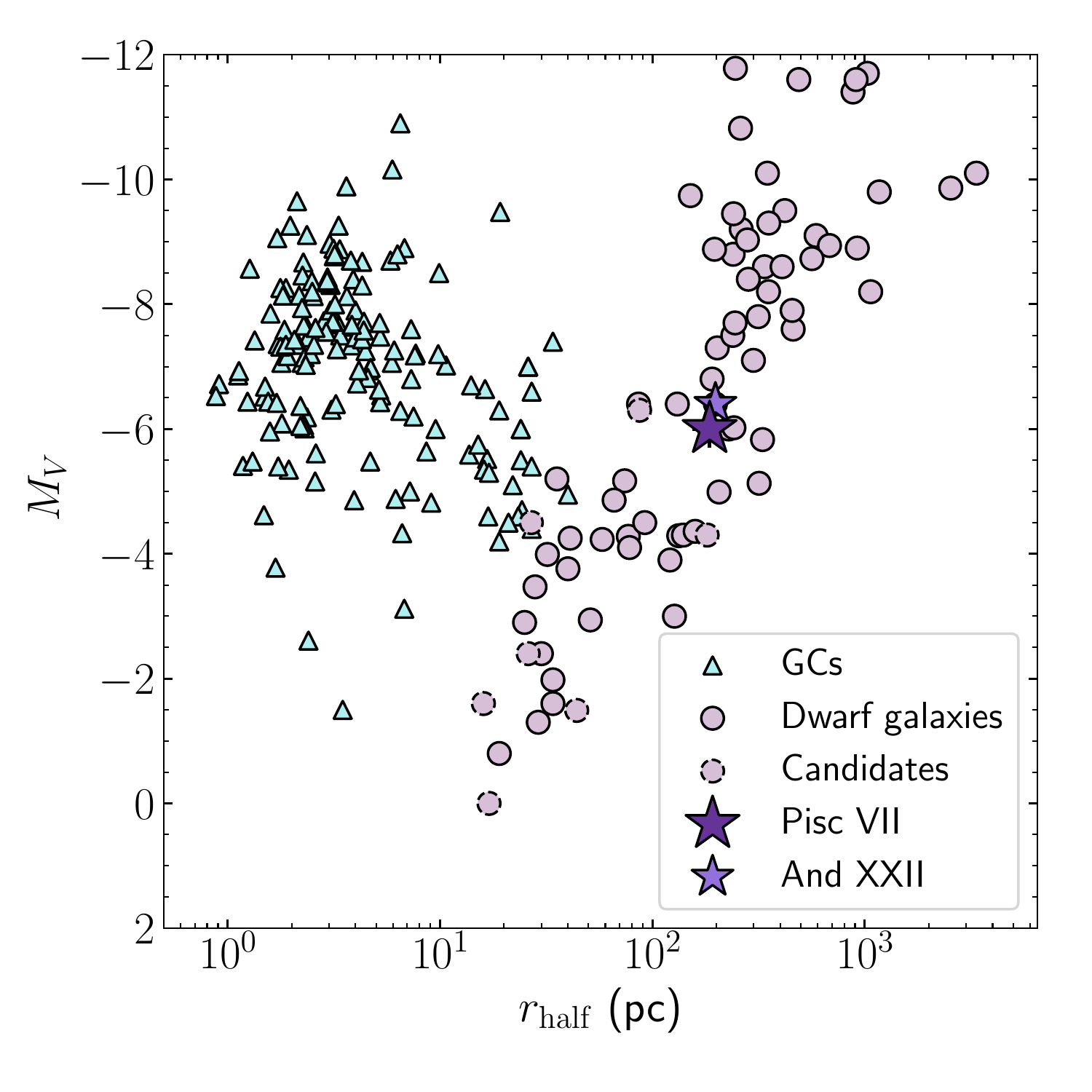}
	\caption{ Above we show the size-luminosity parameter space for dwarf galaxies (light pink circles), dwarf candidates (pink circles with dashed borders) and globular clusters (turquoise triangles). \pvii\ is highlighted as the dark purple star, and M33's other potential satellite -- And XXII -- is shown as a smaller, light purple star. }
    \label{fig:sl}
\end{figure}

The results from this procedure are \Mg\ and \Mr. We can then translate these to an absolute magnitude in the $V$-band of \Mv\ using the colour transforms of \citet{jordi06}.  As \pvii\ is so faint and low density, its stellar populations are uncrowded down to below the HB, so we apply no completeness corrections to these measured values. Our new luminosity is fully consistent with the value we reported in \citet{martinezdelgado22}. Translating this into a luminosity, we measure \Lum. Given the size and luminosity we measure for \pvii, it is a highly probable ultra faint dwarf galaxy rather than a stellar cluster. In Fig.~\ref{fig:sl}, we show the half-light radii for dwarf galaxies (pink circles), candidate dwarf galaxies (pink circles with dashed border) and globular clusters (turquoise triangles) vs. their absolute magnitude \cite[with data taken from][]{harris96,mackey13,koposov15,martin16,martinezvazquez19,simon20,cerny21a,cerny21b,cerny22,cerny23a,savino22}. \pvii\ (dark purple star) is clearly found in the dwarf galaxy regime, and significantly offset from globular clusters of comparable luminosity. To definitively test this, we would require spectroscopy to measure any dark matter contribution and/or a spread in metallicity characteristic of a dwarf galaxy.

\section{Discussion}
\label{sec:discussion}

\begin{figure}
	\includegraphics[width=\columnwidth]{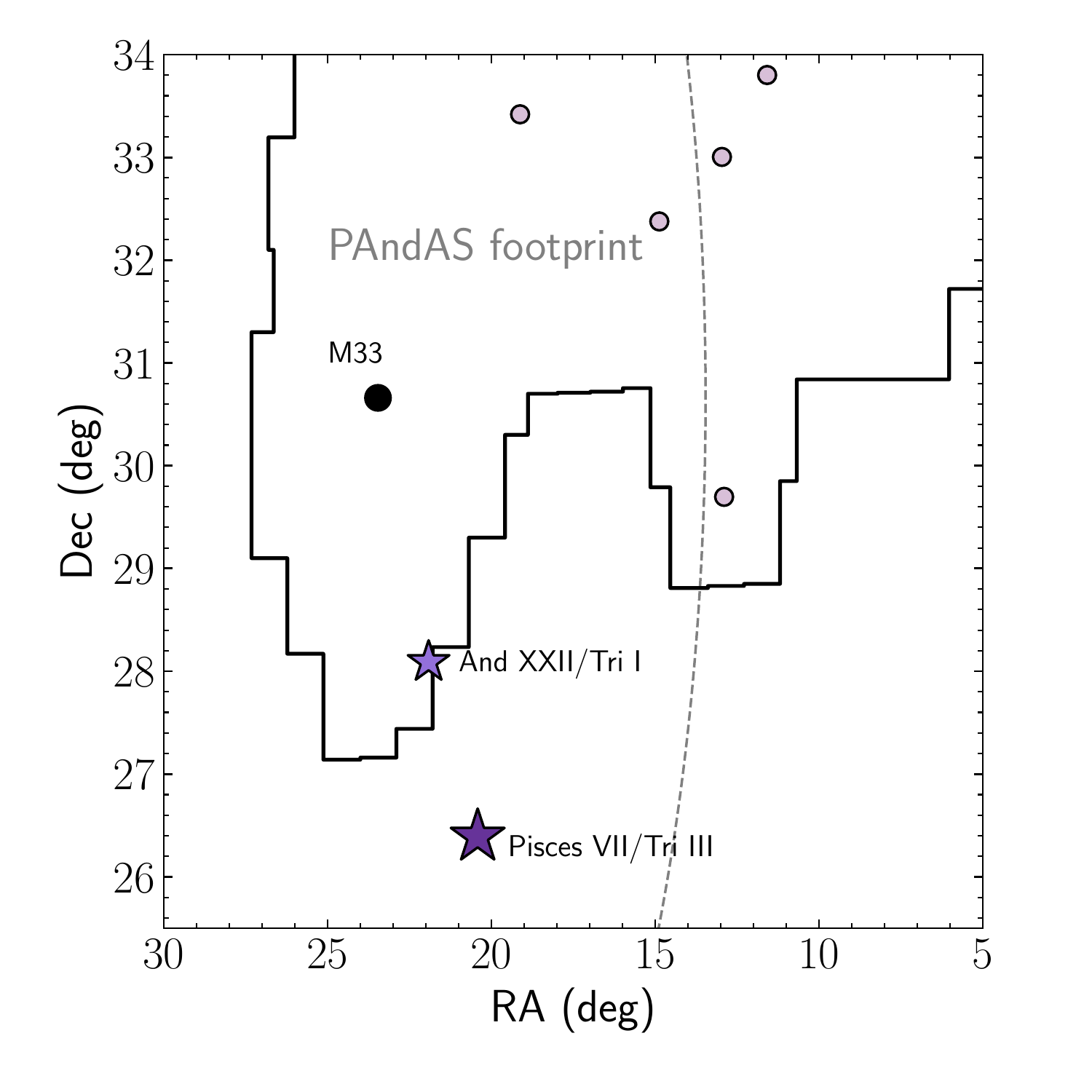}
	\caption{ Here we show the location of \pvii\ (dark purple star) in the Triangulum system. Located outside the PAndAS footprint (shown as the black outline), it is closer in projection and 3D distance to M33 (labelled) than M31. The approximate virial radius of M33 is shown as a dashed gray line. We also show M33's other possible satellite, And XXII, as a light purple star. Interestingly, the two dwarf galaxies appear to be in a direct line from M33.  Other M31 dwarf galaxies in this region are shown as light pink circles.}
    \label{fig:map}
\end{figure}

\subsection{\pvii\ -- a satellite of M31 or M33?}

With our revised distance of \dist, we have shown that \pvii\ is unlikely to be an isolated ultra-faint at the present day. At a distance of \dtri\ from Triangulum, it is comfortably within its predicted dark matter halo \citep{patel18} and therefore a likely M33 satellite. However, with a distance from the centre of M31 of $D_{\rm M31}=285^{+21}_{-18}$~kpc, it could be a far-flung Andromeda satellite. A bulk radial velocity for \pvii\ would likely resolve this ambiguity and be able to determine for sure which of M33 and M31. We visualise the projected location of \pvii\ from both galaxies in Fig.~\ref{fig:map}. 

The proximity of \pvii\ to M33 is interesting, as to-date, M33 has only one known satellite -- And XXII -- which is close to M33 in projection ($\sim43$~kpc). Recent work from \citet{savino22} find it has a 3D separation from M33 of $113^{+31}_{-30}$~kpc and a separation from M31 of $215\pm9$~kpc, making its membership more ambiguous than \pvii. However, based on its radial velocity, \citet{chapman13} argued it is energetically more likely to be a satellite of M33. M33 is predicted to have $8$ satellites with a luminosity equivalent to $10^4\,{\rm L_\odot}$ or greater \citep{dooley17,patel18} within its virial radius. While \pvii\ adds a valuable second data point to M33's satellite system, it retains a missing satellite problem. However, there is cause for optimism. Given that this satellite falls outside the deep PAndAS survey region (which covers only half the virial radius of M33), its existence implies that there may be a wealth of faint dwarf galaxies in the ultra-faint regime awaiting our discovery.

\subsection{Constraining the orbital history of M33}
\label{sec:m33orbit}

If we assume \pvii\ (and And~XXII) are bound to M33, we can further address the issue of M33's recent orbital history. Given its warped HI \cite[e.g.][]{rogstad76} and stellar disks \cite[e.g.][]{mcconnachie09} it has been proposed that M33 has experienced a prior close passage ($<100$~kpc) with M31 a few Gyr ago \citep{mcconnachie09,putman09,teppergarcia20}. However this is in tension with the proper motion informed orbital history of M33 which suggests it is on first infall and that a close passage with M31 in the past 3~Gyr is rare \cite[less than 1\% chance,][]{patel17, vandermarel19}. In this case, the warps must be caused via some other method such as repeated close passages of satellites with M33 \citep{starkenburg16}, asymmetric gas accretion, or long range tidal forces from M31. The first of these suggestions requires M33 to have a significant population of satellites, something which is not yet observed. 

If there truly was a close passage between M31 and M33 a few Gyr ago, much of its halo (and thus satellites) would have been stripped by M31. Any remaining satellites would likely be found close to M33. Given both \pvii\ and And~XXII are found at a large distance from M33 ($>100$~kpc), this would seem to support the model of a first infall rather than a close encounter. In this context, dynamics and proper motions for \pvii\ are very desirable so we could both confirm it is bound to M33, and ascertain whether \pvii\ could have induced warps in M33's disks. Based on our CMD (Fig.~\ref{fig:cmd}) there are a handful of stars in \pvii\ that would be bright enough to obtain spectra for with e.g. the 10~m Keck telescopes. Proper motions can only be acquired with multi-epoch observations from space-based telescopes such as Hubble or JWST. Hopefully, such future observations will allow us to solve these multiple mysteries of M33's evolution.

\subsection{Blue stars in the CMD of \pvii\ -- recent star formation? }

Finally, we briefly discuss the presence of blue stars in the CMD just faint-ward of the HB of \pvii, as well as the potential helium-burning stars just above the HB. The latter indicate a stellar population within the galaxy with an age of $\sim1$~Gyr. Both populations are far more numerous in the dwarf galaxy than in the comparison field (Fig.~\ref{fig:cmd}). They are reminiscent of a similar population seen in shallow HST imaging for the newly discovered isolated UFD, Pegasus~W \citep{mcquinn23}, which are consistent with a young ($\sim0.5-1$~Gyr) populaition. We overlay a younger isochrone in Fig.~\ref{fig:cmd} and find that these stars are consistent with an age of 1.5 Gyr. They could also be blue stragglers, which are main-sequence stars that have undergone a binary merger or mass transfer event which has changed their intrinsic colours, shifting them off the main sequence. However, if they truly are younger stars with ages $\sim$1-2 Gyr, \pvii\ would be the faintest dwarf galaxy to have star formation that has occurred significantly beyond reionisation and would transform our understanding of how UFDs are quenched during this epoch. 

\section{Conclusions}
\label{sec:conclusions}

In this work, we have reassessed the properties of \pvii/Tri III using deep Gemini GMOS-N imaging. We find its structural properties are in good agreement with our previous study from shallower data \citep{martinezdelgado22}. Specifically, we measure its positions as \coords, and a half-light radius of \rhpc. We measure a  slightly lower luminosity of \Lum (\Mv) which is consistent with previous work within $2\sigma$. Based on its size and luminosity, we conclude \pvii\ is a highly probable dwarf galaxy. Our deep data also allows us to determine the distance to \pvii\ from its horizontal branch of \dist. This locates it at a 3D distance of \dtri\, within the halo of M33 making \pvii\ a highly probable satellite of this system. 

\pvii\ is only the second known satellite of M33 and its location at a large distance from M33 may favour a first infall scenario for the orbital history of M33. Proper motions and dynamics would help to determine whether \pvii\ is definitively bound to M33, and thus whether it truly supports the first infall scenario over a prior close encounter with Andromeda. They would also allow us to trace its orbital history to determine whether its past orbit could have induced warps in either its gas or stellar disk. 

The detection of a second far-flung satellite of M33 also supports the existence of a wider satellite system for this galaxy. Beyond the PAndAS footprint, shallow survey data make detections of ultra-faints like \pvii\ extremely challenging. A deep, wide-field survey with an instrument like Hyper-Suprime Cam on the 8-m Subaru telescope that covers the extent of its halo would likely uncover a wealth of substructure in this system, and would definitively assess whether or not M33 has a missing satellite problem.

Finally, we comment on blue stars in the CMD that are consistent with a younger age population (1-2 Gyr). This evidence of recent star formation in such a faint galaxy would be an extremely important finding if verified. Future deep imaging would allow us to study this intriguing population in more detail.

\section*{Acknowledgements}

We thank the anonymous referee for their helpful and constructive comments, which improved the quality of this work. DMD acknowledges financial support from the Talentia Senior Program (through the incentive ASE-136) from Secretar\'\i a General de  Universidades, Investigaci\'{o}n y Tecnolog\'\i a, de la Junta de Andaluc\'\i a. DMD acknowledge funding from the State Agency for Research of the Spanish MCIU through the ``Center of Excellence Severo Ochoa" award to the Instituto de Astrof{\'i}sica de Andaluc{\'i}a (SEV-2017-0709) and project (PDI2020-114581GB-C21/ AEI / 10.13039/501100011033)

M.M. acknowledges support from the Agencia Estatal de Investigaci\'on del Ministerio de Ciencia e Innovaci\'on (MCIN/AEI) under the grant "RR Lyrae stars, a lighthouse to distant galaxies and early galaxy evolution" and the European Regional Development Fun (ERDF) with reference PID2021-127042OB-I00

Based on observations obtained at the international Gemini Observatory, a program of NSF’s NOIRLab, which is managed by the Association of Universities for Research in Astronomy (AURA) under a cooperative agreement with the National Science Foundation on behalf of the Gemini Observatory partnership: the National Science Foundation (United States), National Research Council (Canada), Agencia Nacional de Investigaci\'{o}n y Desarrollo (Chile), Ministerio de Ciencia, Tecnolog\'{i}a e Innovaci\'{o}n (Argentina), Minist\'{e}rio da Ci\^{e}ncia, Tecnologia, Inova\c{c}\~{o}es e Comunica\c{c}\~{o}es (Brazil), and Korea Astronomy and Space Science Institute (Republic of Korea). Processed using the Gemini DRAGONS (Data Reduction for Astronomy from Gemini Observatory North and South) package.

This work was enabled by observations made from the Gemini North telescope, located within the Maunakea Science Reserve and adjacent to the summit of Maunakea. We are grateful for the privilege of observing the Universe from a place that is unique in both its astronomical quality and its cultural significance.

%%%%%%%%%%%%%%%%%%%%%%%%%%%%%%%%%%%%%%%%%%%%%%%%%%
\section*{Data Availability}

Raw images and their associated calibrations are stored on the Gemini archive and are publicly available. Our fully reduced photometric catalogue is available upon reasonable request to the lead author. %{\bf MC: Happy to just make this publicly available. Depends on the feelings of the group.} 



\bibliographystyle{mnras}
\bibliography{dwarf} % if your bibtex file is called example.bib

%\begin{thebibliography}{99}
%\end{thebibliography}

%%%%%%%%%%%%%%%%%%%%%%%%%%%%%%%%%%%%%%%%%%%%%%%%%%

%%%%%%%%%%%%%%%%% APPENDICES %%%%%%%%%%%%%%%%%%%%%

%\appendix


%%%%%%%%%%%%%%%%%%%%%%%%%%%%%%%%%%%%%%%%%%%%%%%%%%

% Don't change these lines
\bsp	% typesetting comment
\label{lastpage}
\end{document}